\documentclass[prb,twocolumn,showpacs,aps,superscriptaddress,floatfix]{revtex4}
\usepackage{amsmath}
\usepackage{amssymb}
\usepackage{bm}
\usepackage{graphicx}
\usepackage{color}
\usepackage{hyperref}
\usepackage{verbatim}% Long comments

\begin{document}

\title{Disorder correction to the N\'{e}el temperature of ruthenium-doped
BaFe$_2$As$_2$: Theoretical analysis}

\author{S.V. Kokanova}
\affiliation{Skolkovo Institute of Science and Technology, Skolkovo
Innovation Center 3, Moscow 143026, Russia}
\affiliation{Moscow Institute of Physics and Technology (State
University), Institutsky lane 9, Dolgoprudny, Moscow region, 141700 Russia}

\author{A.V. Rozhkov}
\affiliation{Skolkovo Institute of Science and Technology, Skolkovo
Innovation Center 3, Moscow 143026, Russia}
\affiliation{Moscow Institute of Physics and Technology (State
University), Institutsky lane 9, Dolgoprudny, Moscow region, 141700 Russia}
\affiliation{Institute for Theoretical and Applied Electrodynamics, Moscow,
125412 Russia}

\begin{abstract}
We analyze theoretically nuclear magnetic resonance data for the
spin-density wave phase in the
ruthenium-doped BaFe$_2$As$_2$.
Since inhomogeneous distribution of Ru~atoms introduces disorder into the
system, experimentally observable random spatial variations of the
spin-density wave order parameter emerge. Using perturbation theory for
the Landau functional, we estimate the disorder-induced correction to the
N\'{e}el temperature for this material. Calculated correction is
significantly smaller than the N\'{e}el temperature itself for all
experimentally relevant doping levels. This implies that, despite
pronounced spatial non-uniformity of the order parameter, the N\'{e}el
temperature is quite insensitive to the disorder created by the dopants.
\end{abstract}

\date{\today}

\maketitle

\section{Introduction}

In this paper we discuss the influence of doping-induced disorder on the
N{\'e}el temperature of the ruthenium-doped 
BaFe$_2$As$_2$.
The compound is a representative of a wide class of pnictide
superconductors, actively studied in the last decade. Many members of this
class, including 
BaFe$_2$As$_2$
itself, experience a transition into a spin-density wave (SDW) phase. The
N{\'e}el temperature for this transition is sensitive to the doping
concentration and decreases monotonically when the doping grows. Beyond
certain doping level
($\sim 25$\%
for Ru-doped
BaFe$_2$As$_2$)
the magnetism is completely replaced by the superconducting phase.
 
%Some doping atoms supress the SDW very efficiently. For example, 7\% of
%cobalt atoms substituting for iron destroy the SDW in BaFe$_2$As$_2$
%completely. Unlike cobalt, ruthenium exerts midler effects on the magnetic
%structure: the low-temperature SDW state can be detected for Ru
%concentration of $20$\% and even slightly higher. This disparity is often
%associated with the fact~\cite{} that ruthenium is isovalent with iron,
%therefore, it does not damage Fermi surface nesting as much as cobalt
%atoms do.  

Although doping by ruthenium atoms is an important experimental 
method~\cite{data_bafeas2010,PhysRevLett.107.267002,
ru_doped_sdw_sc_coex_exper2012,ru_doped_neutron_resonant_mode2013}
to explore electronic correlation effects in
BaFe$_2$As$_2$,
introduction of dopants unavoidably produces crystal
imperfections~\cite{first_princip_ru_doped2013,iron_based2014,
dft_ru_doped2017}.
While disorder might be a
source~\cite{irradiation_disorder2014}
of interesting phenomena, it is often an undesirable factor blurring or
masking an investigated feature. This concern is quite general for doped
iron-based superconductors. Indeed, the presence of imperfections in this
family of superconducting materials is well-documented: inhomogeneities of
the charge density were observed
experimentally~\cite{PSexp1,PSexp2,PSexp3,phasep_bafeas_exp2010,
epl_inhomogen_sc_2011,phasep_exp2012,phasep_exp2016}
and discussed
theoretically~\cite{phasep_pnics_ours2013,hund_phasep_theor2017}
in several publications.

For an imperfect system, it is reasonable to ask to what extent a
particular physical property is affected by the disorder. Depending on the
nature of the physical property under consideration, the answer to this question
may differ. For instance, NMR
measurements~\cite{laplace2012}
for
BaFe$_2$As$_2$
are consistent with the notion that the SDW order parameter varies
markedly over the sample volume. At the same time, our theoretical
analysis of the same data shows that, notwithstanding pronounced
non-uniformity of the ordered state, the N\'{e}el temperature
$T_{\rm N}$
is fairly insensitive to the dopant-induced inhomogeneities.

Our analysis is based on the perturbation
theory~\cite{larkin_ovch_inhomogen_sc1972}
in powers of the disorder strength. A key ingredient of our study is a
phenomenological model for disorder distribution, developed in
Ref.~\onlinecite{laplace2012}
to interpret the NMR data. The correction to the N{\'e}el temperature is
estimated within the Landau functional framework and is determined to be
small. This finding is the main result of our work. It implies that
$T_{\rm N}$
can be reliably calculated, at least in principle, using disorder-free
models, and the N\'{e}el temperature can act as a benchmark characteristic,
useful for checking the validity of theoretical conclusions.

The paper is organized as follows. In 
Sec.~\ref{sec::model}
the model is introduced. The perturbation theory calculations are performed
in
Sec.~\ref{sec::calcs}.
They are applied to the analysis of the data in
Sec.~\ref{sec::data_analysis}.
Section~\ref{sec::conclusions}
contains the discussion and conclusions. Some auxiliary derivations are
presented in Appendix.

\section{Model}
%%%%%%%%%%%%%%%%%%%%%%%%%%%%%%%%%%%%%%%%%%%%%%%%%%
\label{sec::model}
%%%%%%%%%%%%%%%%%%%%%%%%%%%%%%%%%%%%%%%%%%%%%%%%%% 

Our analysis is based on experimental findings of
Ref.~\onlinecite{laplace2012},
which performed NMR studies of the SDW transition in ruthenium-doped
BaFe$_2$As$_2$.
Since ruthenium substitutes iron atoms, chemical formula for the resultant
alloy is
Ba(Fe$_{1-x}$Ru$_x$)$_2$As$_2$,
where the doping concentration $x$ changes in a wide range
$0<x<1$.
Since Ru is isovalent to Fe, it is
believed~\cite{laplace2012}
that doping by ruthenium atoms creates milder modifications to the electron
structure of the compound as compared, for example, with doping by cobalt
atoms. In particular, one may expect that Ru~substitution does not generates
significant de-nesting, since no electrons are introduced due to dopants.
Yet, Ru~doping weakens the SDW phase: the N\'{e}el temperature decreases as
a function of doping, until the SDW is replaced by the superconductivity
above
$\sim 0.3$.

For our study we need to calculate correction to the SDW transition
temperature, which turns out to be small. Consequently, the Landau free
energy functional
\begin{eqnarray}
%%%%%%%%%%%%%%%%%%%%%%%%%%%%%%%%%%%%%%%%%%%%%%%%%%
\label{eq::landau}
%%%%%%%%%%%%%%%%%%%%%%%%%%%%%%%%%%%%%%%%%%%%%%%%%%
F[S({\bf r})]
=
\int 
	\left\{
		C_\| \left[
			\left(\nabla_x S \right)^2
			+
			\left(\nabla_y S \right)^2
		\right]
		+
		C_\perp \left(\nabla_z S \right)^2
	\right.
\\
\nonumber
	\left.
		+
		A S^2
		+
		\frac{B}{2} S^4
	\right\}d^3 {\bf r},
\end{eqnarray} 
describing the material's behavior near the transition, can be justifiably
used. In 
Eq.~(\ref{eq::landau}),
symbol
$S=S({\bf r})$
is the SDW magnetization, which plays the role of the order parameter.
While, in general, the magnetization is a vector, for our purposes it is
sufficient to treat $S$ as a scalar. Coefficients
$C_\|$,
$C_\perp$,
and $B$ are all positive. To account for the quasi-two-dimensional (Q2D)
anisotropy present in the pnictides the coefficients 
$C_\|$
and
$C_\perp$
must satisfy the inequality
$C_\| \gg C_\perp$.
We will make this condition more specific below. 

For the coefficient $A$, we assume that it is spatially inhomogeneous:
$A=A({\bf r})$.
A convenient parametrisation for
$A({\bf r})$
is as follows
\begin{eqnarray}
A({\bf r}) = a [T - T_{\rm N} - \delta T_{\rm N} ({\bf r})],
\quad
a > 0,
\end{eqnarray}
where $T$ is the system temperature, the disorder-averaged N{\'e}el
temperature is
$T_{\rm N}$.
Local variation of the N{\'e}el temperature 
$\delta T_{\rm N} ({\bf r})$
satisfies
\begin{eqnarray}
%%%%%%%%%%%%%%%%%%%%%%%%%%%%%%%%%%%%%%%%%%%%%%%%%%
\label{eq::zero_av}
%%%%%%%%%%%%%%%%%%%%%%%%%%%%%%%%%%%%%%%%%%%%%%%%%% 
\langle \delta T_{\rm N} ({\bf r}) \rangle \equiv 0.
\end{eqnarray} 
Here triangular brackets
$\langle \ldots \rangle$
denote the average over disorder configurations.

Formulating this model, we assumed that the doping-induced disorder affects
the system mostly through the spatial variation of 
$\delta T_{\rm N} ({\bf r})$.
Inhomogeneities of
$C_{\|,\perp}$
and $B$ are much less important, for they contribute to the subleading
corrections. Therefore, we will treat these parameters as if they are
independent of
${\bf r}$.

Variation of $F$ over $S$ gives us the following equation for the order
parameter $S$
\begin{eqnarray}
-\left[
	\xi_\|^2\left(\nabla_x^2 + \nabla_y^2\right)
	+
	\xi_z^2\nabla_z^2
\right] S
- \delta t({\bf r}) S + b S^3 = - t S,
%%%%%%%%%%%%%%%%%%%%%%%%%%%%%%%%%%%%%%%%%%%%%%%%%% 
\label{nonlinear}
%%%%%%%%%%%%%%%%%%%%%%%%%%%%%%%%%%%%%%%%%%%%%%%%%% 
\end{eqnarray}
where coefficient $b$ equals to
$b = B/(a T_{\rm N})$,
the dimensionful parameter
$\xi_\|^2
=
C_\|/\left(aT_{\rm N}\right)$
is the in-plane correlation length,
$\xi_z^2 = C_\perp/\left(aT_{\rm N}\right)$
is the transverse correlation length. Dimensionless variation of the local
N\'{e}el temperature 
$\delta t$
and dimensionless temperature $t$ in
Eq.~(\ref{nonlinear})
are
\begin{eqnarray}
\delta t({\bf r} )
=
\frac{\delta T_{\rm N} \left( {\bf r} \right) }{T_{\rm N}},
\quad
t = \frac{T - T_{\rm N}}{T_{\rm N}}.
\label{eq::5}
\end{eqnarray} 
We want to calculate the lowest (second) order correction
$\Delta T_{\rm N}$
to the N\'{e}el temperature
$T_{\rm N}$
caused by disorder
$\delta T_{\rm N} ({\bf r})$.
Of course, once 
$\Delta T_{\rm N}$
is evaluated, the experimentally measurable transition temperature is
determined as
$T_{\rm N}+ \Delta T_{\rm N}$.
%%%%%%%%%%%%%%%%%%%%%%%%%%%%%%%%%%%%%%%%%%%%%%%%%%
\begin{comment}
\begin{figure}[t]
	\centering{\includegraphics[width=0.99\columnwidth]{dt.eps}}
	\caption{Schematic phase diagram of Ru-doped
BaFe$_2$As$_2$
(superconductivity at higher doping $x$ is not shown).
Theoretically evaluated N\'{e}el temperature 
$T_{\rm N}(x)$,
shown here as black dashed curve, is typically calculated with the help of 
translationally invariant models neglecting any disorder. Disorder,
inevitably present in a doped sample, shifts the theoretical value
$T_{\rm N}(x)$
by finite correction
$\Delta T_{\rm N}(x)$
(the correction is represented by red arrow).
True transition temperature (experimentally measurable quantity)
$T_{\rm N}(x) + \Delta T_{\rm N}(x)$
is depicted by solid blue curve. We argue that 
$\Delta T_{\rm N}$
is small for all experimentally relevant doping concentrations.
%%%%%%%%%%%%%%%%%%%%%%%%%%%%%%%%%%%%%%%%%%%%%%%%%%
\label{fig::dt}
%%%%%%%%%%%%%%%%%%%%%%%%%%%%%%%%%%%%%%%%%%%%%%%%%% 
}
\end{figure}
\end{comment}
%%%%%%%%%%%%%%%%%%%%%%%%%%%%%%%%%%%%%%%%%%%%%%%%%% 
To find
$\Delta T_{\rm N}$
we need to study only the linear part of
Eq.~(\ref{nonlinear}).
Intuitively, one may argue that, since
$O(|S|^3)$
term in
Eq.~(\ref{nonlinear})
is much smaller than 
$O(|S|)$
near the transition temperature,
$O(|S|^3)$
term may be omitted. A more precise line of reasoning is based on the
realization that the transition temperature is controlled by the bilinear
part of the Landau functional: as long as the bilinear form remains
positive-definite, the disordered phase 
($S\equiv 0$)
remains stable. At the transition temperature the lowest eigenvalue of the
bilinear form vanishes (that is, the bilinear form becomes
non-negative-definite). Thus, to calculate the transition temperature, we
need to study the following eigenvalue equation
\begin{eqnarray}
-\left[
	\xi_\|^2\left(\nabla_x^2 + \nabla_y^2\right) + \xi_z^2\nabla_z^2
\right] S - \delta t({\bf r}) S = - t_{\rm N} S.
%%%%%%%%%%%%%%%%%%%%%%%%%%%%%%%%%%%%%%%%%%%%%%%%%% 
\label{eq::sch}
%%%%%%%%%%%%%%%%%%%%%%%%%%%%%%%%%%%%%%%%%%%%%%%%%% 
\end{eqnarray}
Physical meaning of parameter
$t_{\rm N}$
is the dimensionless disorder-induced correction to the transition
temperature:
\begin{eqnarray}
%%%%%%%%%%%%%%%%%%%%%%%%%%%%%%%%%%%%%%%%%%%%%%%%%%
\label{eq:DTN}
%%%%%%%%%%%%%%%%%%%%%%%%%%%%%%%%%%%%%%%%%%%%%%%%%% 
\Delta T_{\rm N} = T_{\rm N} t_{\rm N}.
\end{eqnarray} 
Mathematically, the value of
$-t_{\rm N}$
is the lowest eigenvalue of the linear operator in the left-hand side
of
Eq.~(\ref{eq::sch}).
Once this eigenvalue is known,
relation~(\ref{eq:DTN})
can be used to find the dimensionful correction.

\section{Calculations}
%%%%%%%%%%%%%%%%%%%%%%%%%%%%%%%%%%%%%%%%%%%%%%%%%%
\label{sec::calcs}
%%%%%%%%%%%%%%%%%%%%%%%%%%%%%%%%%%%%%%%%%%%%%%%%%% 

Before starting the calculations of
$t_{\rm N}$,
it is useful to observe that
Eq.~(\ref{eq::sch}) 
is similar to Schr\"odinger equation. This analogy allows us to determine
the correction to the N\'{e}el temperature using a familiar language of the
perturbation theory for a Schr\"odinger operator. Within this analogy, the
quantity
$-\delta t({\bf r})$
plays the role of small perturbation in potential energy, and
$- t_{\rm N}$
is the correction to the lowest eigenvalue of the non-perturbed
Hamiltonian.

For the three-dimensional systems, the perturbative derivation of
$t_{\rm N}$
has been reported in
Ref.~\onlinecite{larkin_ovch_inhomogen_sc1972}.
Since the pnictides are layers systems, they are often described by
two-dimensional or Q2D models. Our main goal in this Section is to adapt
the calculations of
Ref.~\onlinecite{larkin_ovch_inhomogen_sc1972}
to a quasi-two-dimensional system. While our discussion is, in many
respects, similar to the three-dimensional case, yet, certain technical
points require more delicate treatment.

Let us start with the calculations. Using the logic of the perturbation
theory for the Schr{\"o}dinger operator, we will find the correction to the
ground state eigenvalue for the unperturbed operator
\begin{eqnarray}
%%%%%%%%%%%%%%%%%%%%%%%%%%%%%%%%%%%%%%%%%%%%%%%%%%
\label{eq::h0_def}
%%%%%%%%%%%%%%%%%%%%%%%%%%%%%%%%%%%%%%%%%%%%%%%%%% 
H_0 
=
-\left[\xi_\|^2\left(\nabla_x^2 + \nabla_y^2\right) +
\xi_z^2\nabla_z^2\right].
\end{eqnarray}
The unperturbed ground state is equal to
\begin{eqnarray}
	S^{(0)} = \frac{1}{\sqrt{V}},
	\label{S0}
\end{eqnarray}
where $V$ is the volume of the sample. The first-order
correction to order parameter 
$S^{(1)}$
satisfies the equation
\begin{eqnarray}
%%%%%%%%%%%%%%%%%%%%%%%%%%%%%%%%%%%%%%%%%%%%%%%%%%
\label{eq::1st}
%%%%%%%%%%%%%%%%%%%%%%%%%%%%%%%%%%%%%%%%%%%%%%%%%% 
\left[
	\xi_\|^2\left(\nabla_x^2 + \nabla_y^2\right)
	+
	\xi_z^2\nabla_z^2
\right]\! S^{(1)}
+
\frac{\delta t({\bf r})}{\sqrt{V}} 
=
\frac{t^{(1)}_{\rm N}}{\sqrt{V}}.
\end{eqnarray}
Here
$t^{(1)}_{\rm N}$
is the first-order correction to the eigenvalue 
$t_{\rm N}$. 

Averaging this equation over the disorder, we derive, using
Eq.~(\ref{eq::zero_av}),
that
\begin{eqnarray}
%%%%%%%%%%%%%%%%%%%%%%%%%%%%%%%%%%%%%%%%%%%%%%%%%% 
\label{eq::s1_ave}
%%%%%%%%%%%%%%%%%%%%%%%%%%%%%%%%%%%%%%%%%%%%%%%%%% 
\left[
	\xi_\|^2\left(\nabla_x^2 + \nabla_y^2\right) + \xi_z^2\nabla_z^2
\right]
	\langle S^{(1)} \rangle
=
\frac{1}{\sqrt{V}}t^{(1)}_{\rm N}.
\end{eqnarray}
Since
$\langle S^{(1)} \rangle$
is independent of 
${\bf r}$,
we have
$\nabla \langle S^{(1)} \rangle \equiv 0$.
Therefore,
\begin{equation}
t^{(1)}_{\rm N} = 0.
\end{equation}
Substituting this result into
Eq.~(\ref{eq::1st}),
we calculate the first-order correction to order parameter
\begin{eqnarray}	
S^{(1)} ({\bf r})
=
\frac{1}{\sqrt{V}} \int G({\bf r}-{\bf r'})\, \delta t({\bf r'})d^3{\bf r'}.
%%%%%%%%%%%%%%%%%%%%%%%%%%%%%%%%%%%%%%%%%%%%%%%%%% 
\label{S1}
%%%%%%%%%%%%%%%%%%%%%%%%%%%%%%%%%%%%%%%%%%%%%%%%%% 
\end{eqnarray}
In this relation
$G({\bf r})$
is the Green's function of the operator
$H_0$,
Eq.~(\ref{eq::h0_def}):
\begin{eqnarray}
%%%%%%%%%%%%%%%%%%%%%%%%%%%%%%%%%%%%%%%%%%%%%%%%%% 
\label{eq::green_f}
%%%%%%%%%%%%%%%%%%%%%%%%%%%%%%%%%%%%%%%%%%%%%%%%%% 
G({\bf r})
= 
\int\! \frac{d^3{\bf k}}{(2\pi)^3}
	 \frac{e^{i{\bf k}{\bf r}}}
		{\xi_\|^2 \left(k_x^2 + k_y^2\right)
		+
		\xi_z^2 k_z^2}.
%=
%\frac{1}{\xi_\|^2 \xi_z}
%\int\! \frac{d^3{\bf \tilde{k}}}{(2\pi)^3}
%	\frac{e^{i{\bf \tilde{k}}{\bf \tilde{r}}}}
%		{\tilde{k}^2}
%=
%\frac{1}{\xi_\|^2 \xi_z}\frac{1}{4\pi|\bf \tilde{r}|}
%=
%\frac{1}{4\pi\xi_\|^2 \xi_z}
%\frac{1}{\sqrt{\frac{\left(x^2+y^2\right)}{\xi_\|^2} + \frac{z^2}{\xi_z^2}}},
\end{eqnarray}
Fourier transform of
$G({\bf r})$
equals to
\begin{eqnarray}
G_{\bf k} = \frac{1}{\xi_\|^2 \left(k_x^2 + k_y^2\right) + \xi_z^2 k_z^2}.
%%%%%%%%%%%%%%%%%%%%%%%%%%%%%%%%%%%%%%%%%%%%%%%%%% 
\label{GreenFourier}
%%%%%%%%%%%%%%%%%%%%%%%%%%%%%%%%%%%%%%%%%%%%%%%%%% 
\end{eqnarray}

To find the second-order correction to
$t_{\rm N}$
it is necessary to obtain the equation for the second-order correction to
order parameter 
$S^{(2)}$.
Retaining all terms up to the second order in
$\delta t$,
we can write
\begin{eqnarray}
%%%%%%%%%%%%%%%%%%%%%%%%%%%%%%%%%%%%%%%%%%%%%%%%%% 
\label{Second}
%%%%%%%%%%%%%%%%%%%%%%%%%%%%%%%%%%%%%%%%%%%%%%%%%% 
\left[
	\xi_\|^2\left(\nabla_x^2 + \nabla_y^2\right) + \xi_z^2\nabla_z^2
\right]
\left(S^{(0)}+S^{(1)}+S^{(2)}\right)+
\\
\nonumber
+ \delta t \left(S^{(0)}+S^{(1)}\right) = t^{(2)}_{\rm N}S^{(0)}.
\end{eqnarray}
Collecting all second-order terms in this expression, it is possible to
derive for 
$t^{(2)}_{\rm N}$: 
\begin{eqnarray}	
%%%%%%%%%%%%%%%%%%%%%%%%%%%%%%%%%%%%%%%%%%%%%%%%%%
\label{eq::correction}
%%%%%%%%%%%%%%%%%%%%%%%%%%%%%%%%%%%%%%%%%%%%%%%%%%
t^{(2)}_{\rm N}
=
\int\! S^{(0)} \left[
			\xi_\|^2\left(\nabla_x^2 + \nabla_y^2\right)
			+
 			\xi_z^2\nabla_z^2
		\right]S^{(2)}({\bf r}) d^3{\bf r}
+
\\
\nonumber
+ \int\! \delta t({\bf r})S^{(0)}S^{(1)}({\bf r}) d^3{\bf r}. 
\end{eqnarray}
Since
$S^{(0)}$
is independent of
${\bf r}$,
the first term can be written as a divergence of some vector field.
Therefore, the volume integral can be replaced with a surface integral,
which vanishes for periodic boundary conditions. Thus
\begin{eqnarray}	
%%%%%%%%%%%%%%%%%%%%%%%%%%%%%%%%%%%%%%%%%%%%%%%%%% 
\label{eq::correction0}
%%%%%%%%%%%%%%%%%%%%%%%%%%%%%%%%%%%%%%%%%%%%%%%%%% 
t^{(2)}_{\rm N}
&=&
\int\! \delta t({\bf r})S^{(0)}S^{(1)}({\bf r}) d^3{\bf r}
\\
\nonumber 
&=&
\frac{1}{V}\int\! \delta t({\bf r}) \delta t({\bf r}') G({\bf r}-{\bf r}')
d^3{\bf r} \, d^3{\bf r}'.
\end{eqnarray}
This equation explicitly demonstrates that the correction
$t^{(2)}_{\rm N}$
is a random quantity, a (bilinear) functional of the disorder
configuration
$\delta t$.
However, we prove in 
Appendix~\ref{app::A}
that the dispersion of
$t^{(2)}_{\rm N}$
vanishes in the thermodynamic limit. Thus, since
$\langle t^{(2)}_{\rm N}\rangle
\approx
t^{(2)}_{\rm N}$,
it is permissible to work with the average value of
$t^{(2)}_{\rm N}$.
Once the disorder averaging in
Eq.~(\ref{eq::correction})
is performed, we obtain
\begin{eqnarray}	
%%%%%%%%%%%%%%%%%%%%%%%%%%%%%%%%%%%%%%%%%%%%%%%%%% 
\label{AverageCorrection}
%%%%%%%%%%%%%%%%%%%%%%%%%%%%%%%%%%%%%%%%%%%%%%%%%% 
\langle t^{(2)}_{\rm N}\rangle
&=&
\frac{1}{V} \int\! \langle \delta t({\bf r})\, \delta t({\bf r}')\rangle
G({\bf r}-{\bf r'})\, d^3{\bf r}\, d^3{\bf r}'
\\
\nonumber 
&=&
\int\! \tau ( {\bf r} ) G( {\bf r} )\, d^3{\bf r}.
\end{eqnarray}

Below we will assume that the disorder correlation function 
\begin{eqnarray}
\tau({\bf r}-{\bf r'})
=
\langle \delta t({\bf r})\, \delta t({\bf r}')\rangle
%%%%%%%%%%%%%%%%%%%%%%%%%%%%%%%%%%%%%%%%%%%%%%%%%% 
\label{CorrelationFunction}
%%%%%%%%%%%%%%%%%%%%%%%%%%%%%%%%%%%%%%%%%%%%%%%%%% 
\end{eqnarray}
has the following structure
\begin{eqnarray}
%%%%%%%%%%%%%%%%%%%%%%%%%%%%%%%%%%%%%%%%%%%%%%%%%%
\label{eq::corr_func}
%%%%%%%%%%%%%%%%%%%%%%%%%%%%%%%%%%%%%%%%%%%%%%%%%% 
\tau({\bf r})
=
\langle \Delta t^2\rangle
\exp\left(-\frac{x^2+y^2}{2r_0^2}\right)\delta\left(\frac{z}{s}\right).
\end{eqnarray}	
In this expression,
$\langle \Delta t^2\rangle$ 
is the variance of the local dimensionless N\'{e}el temperature, $s$ is the
distance between Fe layers, 
$r_0$ 
is the disorder correlation length in a single Fe layer. The distribution
of the Ru atoms in neighboring layers is assumed to be uncorrelated. This
feature is captured by 
$\delta (z/s)$
in
Eq.~(\ref{eq::corr_func}).

Switching in 
Eq.~(\ref{AverageCorrection})
from integration over real space to integration over momentum space, we
find~\cite{larkin_ovch_inhomogen_sc1972}
\begin{eqnarray}
\langle t^{(2)}_{\rm N}\rangle
=
\int\! \frac{d^3{\bf k}}{(2\pi)^3} \tau_{\bf k}G_{\bf k},
%%%%%%%%%%%%%%%%%%%%%%%%%%%%%%%%%%%%%%%%%%%%%%%%%%
\label{Correctionk}
%%%%%%%%%%%%%%%%%%%%%%%%%%%%%%%%%%%%%%%%%%%%%%%%%%
\end{eqnarray}
where the Fourier transform of the correlation function
$\tau({\bf r})$
is
\begin{eqnarray}
\tau_{\bf k}
=
2\pi \langle \Delta t^2\rangle r_0^2 s\,
	\exp\!\left[-\frac{r_0^2}{2}\left(k_x^2 + k_y^2\right)\right].
%%%%%%%%%%%%%%%%%%%%%%%%%%%%%%%%%%%%%%%%%%%%%%%%%% 
\label{CorrelFourier}	
%%%%%%%%%%%%%%%%%%%%%%%%%%%%%%%%%%%%%%%%%%%%%%%%%% 
\end{eqnarray}
Equation~(\ref{Correctionk}),
with the help of
Eqs.~(\ref{GreenFourier})
and~(\ref{CorrelFourier}),
can be re-written as
\begin{eqnarray}
\langle t^{(2)}_{\rm N}\rangle
= 
\frac{\langle \Delta t^2\rangle r_0^2 s}{(2\pi)^2}
\int\! d^3{\bf k} 
	\frac{\exp\left(-\left(k_x^2 + k_y^2\right)r_0^2/2\right)}
	{\xi_\|^2\left(k_x^2 + k_y^2\right) + \xi_z^2 k_z^2}.
\end{eqnarray}
Here the integration over
$k_x$
and
$k_y$
is performed from $-\infty$ to $\infty$. At the same time, the integration
over
$k_z$
is from
$-\pi/s$
to
$\pi/s$.
Taking this into account we obtain
\begin{eqnarray}
%%%%%%%%%%%%%%%%%%%%%%%%%%%%%%%%%%%%%%%%%%%%%%%%%%
\label{eq::t2_integr}
%%%%%%%%%%%%%%%%%%%%%%%%%%%%%%%%%%%%%%%%%%%%%%%%%% 
 \! \left<t^{(2)}_{\rm N}\right>
=
\frac{\langle \Delta t^2\rangle r_0^2 s}{4\pi} \!
\int\limits_0^\infty\! dk_\|^2 
\int\limits_{-\pi/s}^{\pi/s}\! dk_z 
	\frac{\exp\left(-k_\|^2r_0^2/2 \right)}
	{\xi_\|^2 k_\|^2 + \xi_z^2 k_z^2},
\end{eqnarray}
where
$k_\|^2 = k_x^2+k_y^2$. 
It is important to note that the correction is infinite for two-dimensional
systems. Indeed, the integral in
Eq.~(\ref{eq::t2_integr}) 
diverges logarithmically in the limit
$\xi_z = 0$.
To regularize the integral we evaluate it at finite
$\xi_z$,
that is, in Q2D setting. First of all, we denote
$q = k_\|^2r_0^2$,
$\alpha = \xi_z/s$,
$\beta = \xi_\|/r_0$,
$k_z s = u$
and integrate last equation over $q$ by parts:
\begin{eqnarray}
%%%%%%%%%%%%%%%%%%%%%%%%%%%%%%%%%%%%%%%%%%%%%%%%%%
\label{eq::integ_eval}
%%%%%%%%%%%%%%%%%%%%%%%%%%%%%%%%%%%%%%%%%%%%%%%%%% 
 \! \langle
	t^{(2)}_{\rm N}\rangle
=
\frac{\langle \Delta t^2\rangle}{4\pi \beta^2} \!
\int_{-\pi}^{\pi} du \int_0^\infty\!  dq 
	\frac{\exp\left(-\frac{q}{2} \right)}
		{\left[q + \alpha^2 u^2/\beta^2\right]}
=
\\
\nonumber
=
-\frac{\langle \Delta t^2\rangle }{4\pi \beta^2} \!
\int_{-\pi}^{\pi} du \ln\left(0.89\frac{\alpha^2 u^2}{\beta^2}\right)
	\left[1 + O\left(\frac{\alpha^2 u^2}{\beta^2}\right)\right].
\end{eqnarray}
Here we assume that
$\alpha^2 \pi^2/\beta^2 \ll 1$.
This condition will be discussed in
subsection~\ref{subsect::anisotropy}.

Returning to the evaluation of 
$\langle t^{(2)}_{\rm N} \rangle$,
we perform the integration over $u$:
\begin{eqnarray}
%%%%%%%%%%%%%%%%%%%%%%%%%%%%%%%%%%%%%%%%%%%%%%%%%%
\label{eq::integ_final}
%%%%%%%%%%%%%%%%%%%%%%%%%%%%%%%%%%%%%%%%%%%%%%%%%% 
\langle
	t^{(2)}_{\rm N}
\rangle
\approx
\frac{\langle \Delta t^2\rangle}{\beta^2}
\left[1-\ln\left(0.94 \frac{\pi \alpha}{\beta}\right)\right].
%%%%%%%%%%%%%%%%%%%%%%%%%%%%%%%%%%%%%%%%%%%%%%%%%% 
\label{log} 
%%%%%%%%%%%%%%%%%%%%%%%%%%%%%%%%%%%%%%%%%%%%%%%%%% 
\end{eqnarray}
For the logarithmic function in this expression we expect, as usual, that
its value is of the order of unity. Thus
\begin{eqnarray}
%%%%%%%%%%%%%%%%%%%%%%%%%%%%%%%%%%%%%%%%%%%%%%%%%%
\label{eq::FinalCorrection}
%%%%%%%%%%%%%%%%%%%%%%%%%%%%%%%%%%%%%%%%%%%%%%%%%%
\langle t^{(2)}_{\rm N}\rangle
\sim
\frac{\langle \Delta t^2\rangle }{\beta^2}
= \frac{r_0^2 }{\xi_\|^2}\langle \Delta t^2\rangle.
\end{eqnarray}
Thus, the second-order correction to dimensionless N\'{e}el
temperature~(\ref{eq::FinalCorrection})
depends on the variance of the local dimensionless N\'{e}el temperature and
on the ratio of the in-plane length
$\xi_\|$
and disorder correlation length in a single Fe layer
$r_0$.
As for the interlayer parameters $s$ and
$\xi_z$,
they introduce weak logarithmic correction to the main result. This
correction was neglected in
Eq.~(\ref{eq::FinalCorrection}).

Obviously, 
Eq.~(\ref{eq::FinalCorrection})
is applicable not only for antiferromagnets, but also for superconductors,
as well as other ordered phases. As a specific application, in the next
section we will use this formula to find the corrections to the N\'{e}el
temperature of doped
BaFe$_2$As$_2$.

\section{Analysis of experimental data}
%%%%%%%%%%%%%%%%%%%%%%%%%%%%%%%%%%%%%%%%%%%%%%%%%%
\label{sec::data_analysis}
%%%%%%%%%%%%%%%%%%%%%%%%%%%%%%%%%%%%%%%%%%%%%%%%%% 

In this section we will apply 
Eq.~(\ref{eq::FinalCorrection})
to the analysis of the data published in
Ref.~\onlinecite{laplace2012}.
This paper is of particular interest for us here, since it discusses
statistical properties of the local N\'{e}el temperature $\delta T_{\rm N}
({\bf r})$
for doped
BaFe$_2$As$_2$.
Namely, the authors of
Ref.~\onlinecite{laplace2012}
have concluded that their data is consistent with the assumption that 
$\delta T_{\rm N} ({\bf r})$
is obtained by coarse-graining of the random dopant distribution over 
small, but finite, patches of the underlying two-dimensional lattice. The
model for
$\delta T_{\rm N} ({\bf r})$
is formulated~\cite{laplace2012}
in the following manner. Initially, the whole two-dimensional lattice of a
Fe layer is split into square patches. The size of each patch is
$4\times4$
unit cells (obviously, a patch contains
$N=16$
unit cells). For a particular distribution of Ru atoms over a layer, one
defines a function
$n ({\bf r})$,
which is a number of Ru atoms within a patch located at
${\bf r}$.
As a result, the local coarse-grained doping
$x_{\rm loc}({\bf r}) = n({\bf r})/N$
is introduced. Disorder-average of this function is equal to the average
doping:
\begin{eqnarray}
%%%%%%%%%%%%%%%%%%%%%%%%%%%%%%%%%%%%%%%%%%%%%%%%%%
\label{eq::x_av}
%%%%%%%%%%%%%%%%%%%%%%%%%%%%%%%%%%%%%%%%%%%%%%%%%% 
\langle x_{\rm loc} ({\bf r}) \rangle = x.
\end{eqnarray} 
Function
$x_{\rm loc}({\bf r})$
is used to determine local variation of the N\'{e}el temperature according
to the rule
\begin{eqnarray} 
%%%%%%%%%%%%%%%%%%%%%%%%%%%%%%%%%%%%%%%%%%%%%%%%%%
\label{eq::Tvsx}
%%%%%%%%%%%%%%%%%%%%%%%%%%%%%%%%%%%%%%%%%%%%%%%%%% 
\delta T_{\rm N} ({\bf r}) = T_{\rm N} (x_{\rm loc} ({\bf r}))
-
\langle T_{\rm N} (x_{\rm loc} ({\bf r})) \rangle,
\end{eqnarray} 
where the dependence of the N\'{e}el temperature
$T_{\rm N} =  T_{\rm N} (x)$
on the average doping $x$ is directly measured experimentally. We find that
the linear fit
\begin{eqnarray}
%%%%%%%%%%%%%%%%%%%%%%%%%%%%%%%%%%%%%%%%%%%%%%%%%%
\label{eq::fit}
%%%%%%%%%%%%%%%%%%%%%%%%%%%%%%%%%%%%%%%%%%%%%%%%%% 
T_{\rm N} (x) = T_{\rm N}^0(1-\gamma x),
\text{\ where\ }
T_{\rm N}^0 = 140\,{\rm K},\  \gamma = 2,
\end{eqnarray}
accurately describes the data.
Table~\ref{table}
attests to the quality of this approximation.
Formula~(\ref{eq::fit})
works well in the interval
$0<x<0.2$.
For larger doping levels, the N\'{e}el temperature is expected to decrease
faster than described by 
Eq.~(\ref{eq::fit}).
(Therefore, our results will gradually become less accurate when $x$
grows beyond 0.2.)

The outlined disorder model allows us to obtain both
$r_0$
and
$\langle \Delta t^2 \rangle$
for our 
Eq.~(\ref{eq::FinalCorrection}).
Mathematically speaking, the patch size corresponds to the correlation length 
$r_0$
in
Eq.~(\ref{eq::corr_func}).
Indeed, if
$|{\bf r} - {\bf r}'| > r_0$,
the random quantities
$\delta T_{\rm N} ({\bf r})$
and
$\delta T_{\rm N} ({\bf r}')$
characterize different patches. Consequently, they are uncorrelated, which
means that
$\langle t({\bf r})t({\bf r}') \rangle \approx 0$,
in agreement with
Eq.~(\ref{eq::corr_func}).
Therefore, we can write
\begin{eqnarray}
%%%%%%%%%%%%%%%%%%%%%%%%%%%%%%%%%%%%%%%%%%%%%%%%%%
\label{eq::r0_estimate}
%%%%%%%%%%%%%%%%%%%%%%%%%%%%%%%%%%%%%%%%%%%%%%%%%% 
r_0 \approx 4a_0 \approx 11\,{\rm \AA},
\end{eqnarray} 
where
$a_0\approx 2.8$\,\AA\/ 
is the unit cell size, see Ref.~\onlinecite{data_bafeas2010}. Consistent with
Ref.~\onlinecite{laplace2012},
we assume that the in-layer cell is defined in such a manner that there is
one iron atom per cell. 

\begin{table}[t!]
  \begin{center}
    \begin{tabular}{||c||c|c|c|c|c||} % <-- Alignments: 1st column left, 2nd middle
\hline
          \  & $x=0.00$ & $x=0.05$ & $x=0.15$ & $x=0.20$ & $x=0.25$ \\
      \hline\hline
      $T_{\rm N} (x)$, K, & 135 & 130 & 95 & 80 & 55 \\
      Ref.~\onlinecite{laplace2012}&  &  &  &  & \\
      \hline
      $T_{\rm N} (x)$, K, & 140 & 126 & 98 & 84 & 70 \\
       Eq (\ref{eq::fit}) &  &  &  &  &  \\
      \hline\hline
      $\Delta T_{\rm N} (x)$, K, & 0 & 3.3 & 6.9 & 7.4 & 7.2 \\
       Eq (\ref{eq::num_correction}) &  &  &  &  &  \\
\hline
    \end{tabular}
  \end{center}
\caption{Experimentally measured N\'{e}el temperature (from
Ref.~\onlinecite{laplace2012}), 
analytical fit to this temperature, see
Eq.~(\ref{eq::fit}),
and disorder-induced correction to the transition temperature,
Eq.~(\ref{eq::num_correction}),
for several doping concentrations $x$. The values of $x$ from this table
correspond to the experimentally studied samples,
Ref.~\onlinecite{laplace2012}.
The presented data shows that the fit (\ref{eq::fit}) works well for
$x \leq 0.2$.
Only when
$x=0.25$,
significant discrepancy between the measured and fitted N\'{e}el
temperatures emerges. Correction
$\Delta T_{\rm N}$
remains small (ratio
$\Delta T_{\rm N}/ T_{\rm N}$
remains of the order of 10\% or less) for all $x$.
%%%%%%%%%%%%%%%%%%%%%%%%%%%%%%%%%%%%%%%%%%%%%%%%%% 
\label{table}
%%%%%%%%%%%%%%%%%%%%%%%%%%%%%%%%%%%%%%%%%%%%%%%%%% 
}
\end{table}
\begin{figure}[t]
	\centering{\includegraphics[width=0.48\textwidth]{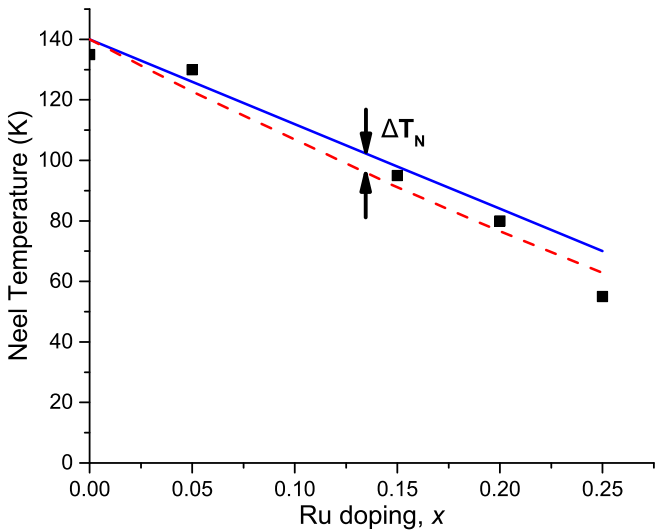}}
	\caption{N{\'e}el temperature for Ru-doped
BaFe$_2$As$_2$
versus doping $x$. Experimental values of the N\'{e}el temperature
$T_{\rm N}(x)$
are shown here as black squares. The data is approximated by solid (blue)
straight line, see
Eq.~(\ref{eq::fit}).
The N\'{e}el temperature for a doped sample without any disorder is depicted by
dashed (red) curve. The disorder-induced correction to the N{\'e}el temperature is 
represented by the difference between the solid and dashed curves. It is estimated
using
Eq.~(\ref{eq::num_correction}). 
%%%%%%%%%%%%%%%%%%%%%%%%%%%%%%%%%%%%%%%%%%%%%%%%%%
\label{fig::T}
%%%%%%%%%%%%%%%%%%%%%%%%%%%%%%%%%%%%%%%%%%%%%%%%%% 
}
\end{figure}

Quantity
$\langle \Delta t^2 \rangle = \langle [\delta t({\bf r})]^2 \rangle$
is the variance of 
$\delta t$
within a single patch. It can be calculated as follows. Combining Eq.~(\ref{eq::5})
Eq.~(\ref{eq::Tvsx})
and~(\ref{eq::fit})
we write:
\begin{eqnarray}
\delta t ({\bf r})
=
\frac{N-\gamma n({\bf r})}{N-\gamma n } - 1
=
\frac{\gamma \left( n - n({\bf r})\right)}{N-\gamma n },
\end{eqnarray}
where
$n = N x$
is the average number of impurities in a patch. The variance of
$\delta t$
within a single patch is
\begin{eqnarray} 
%%%%%%%%%%%%%%%%%%%%%%%%%%%%%%%%%%%%%%%%%%%%%%%%%%
\label{eq::Dt2}
%%%%%%%%%%%%%%%%%%%%%%%%%%%%%%%%%%%%%%%%%%%%%%%%%% 
\langle \Delta t^2\rangle
=
\frac{\gamma^2\langle \Delta n^2\rangle}{\left( N - \gamma n\right)^2},
\quad
\text{where}
\quad
\Delta n = n ({\bf r}) - n.
\end{eqnarray} 
This equation reduces the task of calculating 
$\langle \Delta t^2\rangle$
to the calculation of 
$\langle \Delta n^2\rangle$.
(The latter average will be fond below with the help of the outlined
disorder model.) Such a simple relation between the two is a consequence of
linear dependence of 
$T_{\rm N}$
on doping $x$. In principle, to improve agreement with the experimental
data, one can introduce non-linear terms to
Eq.~(\ref{eq::fit}).
However, we expect that this modification does not significantly change
final results, but only complicates calculations. To keep our formalism
simple and intuitively clear, we always use
Eq.~(\ref{eq::fit}) 
in our study.

Quantity 
$\langle \Delta n^2\rangle$
characterizes the distribution of impurities within a single patch. It can
be easily estimated as follows. The number of impurities in a patch
$n ({\bf r})$
is a random quantity with the binomial distribution. Its number of
``attempts" coincides with the number of unit cells inside a patch:
$N=16$.
The distribution is characterized by ``success probability"
$p=n/N = x$
(this is the probability of finding a Ru atom at a given unit cell inside
the patch). For the binomial distribution with these parameters the answer
is 
$\langle \Delta n^2\rangle = N p (1-p) = n (N-n)/N$.
Substituting this relation into
Eq.~(\ref{eq::Dt2}),
we find 
\begin{eqnarray}
	\langle \Delta t^2\rangle
=
\frac{\gamma^2 n (N-n)}{N\left( N - \gamma n\right)^2}.
%%%%%%%%%%%%%%%%%%%%%%%%%%%%%%%%%%%%%%%%%%%%%%%%%% 
\label{eq::variance_t}
%%%%%%%%%%%%%%%%%%%%%%%%%%%%%%%%%%%%%%%%%%%%%%%%%% 
\end{eqnarray}

Using
Eq.~(\ref{eq::FinalCorrection})
and
Eq.~(\ref{eq::variance_t})
we determine the correction to dimensionless N\'{e}el temperature
\begin{eqnarray}
%%%%%%%%%%%%%%%%%%%%%%%%%%%%%%%%%%%%%%%%%%%%%%%%%% 
\label{FinalEq}
%%%%%%%%%%%%%%%%%%%%%%%%%%%%%%%%%%%%%%%%%%%%%%%%%% 
\langle t^{(2)}_{\rm N}\rangle
\thickapprox 
\frac{r_0^2 \gamma^2 n (N-n )}{N\xi_{\|}^2 \left(N -  \gamma n \right)^2}.
\end{eqnarray}
The dimensionful correction to N\'{e}el temperature can be found with the help of
Eq.~(\ref{eq:DTN}) 
\begin{eqnarray}
\Delta T_{\rm N}
=
\langle t^{(2)}_{\rm N}\rangle T_{\rm N}^0 \left(1 - \gamma \frac{n}{N}\right)
\thickapprox 
\frac{r_0^2 \gamma^2 n (N-n )}{\xi_{\|}^2 N^2\left(N -  \gamma n\right)}T_{\rm N}^0,
%%%%%%%%%%%%%%%%%%%%%%%%%%%%%%%%%%%%%%%%%%%%%%%%%% 
\label{FinalEqKelv}
%%%%%%%%%%%%%%%%%%%%%%%%%%%%%%%%%%%%%%%%%%%%%%%%%% 
\end{eqnarray}
or, with estimate Eq.~(\ref{eq::r0_estimate}), it is equivalent to
\begin{eqnarray} 
\Delta T_{\rm N}
\thickapprox 
\frac{a_0^2 \gamma^2 x (1-x )}{\xi_{\|}^2 \left(1 -  \gamma x\right)}T_{\rm N}^0.
\label{eq::final_correction_with_xi}
\end{eqnarray}

To calculate this correction, the last quantity we need to find is
$\xi_{\|}$.
We estimate 
$\xi_\|$
from the microscopic BCS-like theory:
$\xi_\| \approx 0.13 {v_{\rm F}}/{T_{\rm N}}$,
where
$v_{\rm F}$
is the Fermi velocity. This equation is valid provided that the system's
ordered phase is of mean field BCS character. For more details, one may
consult standard textbook, such as
Refs.~\onlinecite{deGennes,geballe1984}.
Thus, we can write
\begin{eqnarray}
%%%%%%%%%%%%%%%%%%%%%%%%%%%%%%%%%%%%%%%%%%%%%%%%%%
\label{eq::xi_bcs}
%%%%%%%%%%%%%%%%%%%%%%%%%%%%%%%%%%%%%%%%%%%%%%%%%% 
\xi_\| (x) \approx \frac{0.13\, v_{\rm F}}{T_{\rm N}^0 (1 - \gamma x)},
\end{eqnarray} 
where experimentally
measured~\cite{PhysRevLett.107.267002}
value of the Fermi velocity for the compound is virtually
doping-independent and equal to
$v_{\rm F} \sim 0.7$\,eV\AA.

Finally, combining
Eqs.~(\ref{eq::final_correction_with_xi})
and~(\ref{eq::xi_bcs}),
we find 
\begin{eqnarray}
\Delta T_{\rm N}
\thickapprox 
\frac{a_0^2 \gamma^2 x (1-x )\left(1 -  \gamma x\right)}{(0.13 v_{\rm F})^2}(T_{\rm N}^0)^3.
\end{eqnarray}
Once all constants are substituted, equation for the correction to the
N\'{e}el temperature reads:
\begin{eqnarray}
%%%%%%%%%%%%%%%%%%%%%%%%%%%%%%%%%%%%%%%%%%%%%%%%%
\label{eq::num_correction}
%%%%%%%%%%%%%%%%%%%%%%%%%%%%%%%%%%%%%%%%%%%%%%%%%
\Delta T_{\rm N}
\thickapprox 
77 x (1-x )\left(1 -  2 x\right).
\end{eqnarray}

This correction is calculated for several concentrations of Ru~atoms, see
Table~\ref{table}.
The values of $x$ from the Table correspond to the doping levels of the
samples studied in
Ref.~\onlinecite{laplace2012}.
We see that the correction is quite small for
$x < 0.2$.
Figure~\ref{fig::T}
offers additional illustration to this conclusion. Beyond doping
$\sim 0.2$
the system quickly becomes superconducting. Thus, for most of the doping
range where the SDW exists, the disorder-induced corrections to the
N\'{e}el temperature remain weak.

\section{Discussion}
%%%%%%%%%%%%%%%%%%%%%%%%%%%%%%%%%%%%%%%%%%%%%%%%%% 
\label{sec::conclusions}
%%%%%%%%%%%%%%%%%%%%%%%%%%%%%%%%%%%%%%%%%%%%%%%%%%

\subsection{Relevance for other compounds}

The presented calculations are simple and intuitively clear. They also
convey a useful piece of information about
BaFe$_2$As$_2$.
One might inquire if other compounds can be analyzed in a similar manner.

Our procedure depends crucially on the fact that Ru atoms are isovalent
with iron atoms which Ru atoms substitute. Consequently, the doping
does not introduce significant modifications to the electronic structure of
the 
material~\cite{PhysRevLett.107.267002}.
This allows us to write the simplest Landau 
functional~(\ref{eq::landau}),
which remains applicable as long as the Fermi surface nesting is
maintained. For different choice of the dopants, the doping may act to
erode nesting, causing significant modifications to the structure of 
functional~(\ref{eq::landau}).
In this situation, it becomes difficult to justify our model in its present
form. Thus, we must limit ourselves by materials with isovalent doping.

For isovalent doping, 
Eq.~(\ref{eq::FinalCorrection})
is valid, and can be used to estimate the disorder-induced correction. This
equation, of course, requires a practical model of disorder in the
material. The model cannot be obtained by theoretical means, and must be
supplied by experiment. For the very least, parameters
$r_0$
and
$\langle \Delta t^2 \rangle$
must be known. Obviously, 
Ref.~\onlinecite{laplace2012}
fulfills these requirements for
BaFe$_2$As$_2$.
Execution of the similar experimental studies to other pnictide
superconductors may bring useful results about the role of the disorder in
these materials.

\subsection{Comparison of $\xi_\|$ and $r_0$}
%%%%%%%%%%%%%%%%%%%%%%%%%%%%%%%%%%%%%%%%%%%%%%%%%%
\label{subsec::xi}
%%%%%%%%%%%%%%%%%%%%%%%%%%%%%%%%%%%%%%%%%%%%%%%%%% 

Equation~(\ref{eq::xi_bcs})
allows us to estimate 
$\xi_{\|}$
for different values of $x$.
We determine that 
$\xi_\|$
varies between
$7.5$\,\AA\ 
at
$x=0$
to
$15$\,\AA\ 
at
$x=0.25$.
The disorder correlation length
$r_0$,
introduced and discussed in
Sections~\ref{sec::calcs}
and~\ref{sec::data_analysis},
is of the same order:
\begin{eqnarray}
%%%%%%%%%%%%%%%%%%%%%%%%%%%%%%%%%%%%%%%%%%%%%%%%%%
\label{eq::xi_estimate}
%%%%%%%%%%%%%%%%%%%%%%%%%%%%%%%%%%%%%%%%%%%%%%%%%%
\xi_\| \sim r_0,
\end{eqnarray}
see 
estimate~(\ref{eq::r0_estimate}).
This relation is not a coincidence, and can be explained as follows. Purely
local
functional~(\ref{eq::landau})
is an approximation to a more complicated functional with a non-local
kernel. The kernel is spread over a finite size, which we denote
$\xi_\|$.
(When the system obeys the BCS theory, the kernel may be explicitly
evaluated, see, for example, Chapter~7 of the de~Gennes
book~\cite{deGennes}.
However, we expect that the non-local functional itself, as well as the
scale
$\xi_\|$,
are well-defined concepts, even when the BCS microscopic theory is
inapplicable.) To determine the free energy density at a given point
${\bf R}$,
such a functional averages the system's properties over a circle of radius
$\xi_\|$
centered at
${\bf R}$.
For smooth variation of 
$S({\bf r})$,
the non-local functional may be replaced by purely local
Eq.~(\ref{eq::landau}).
Within this simplified formalism, the length scale
$\xi_\|$
emerges as a coefficient in front of the derivatives in
Eq.~(\ref{nonlinear}).
This argument implies that the parameter
$\xi_\|$
describes the smallest length scale below which the
functional~(\ref{eq::landau})
is undefined, and any fragment of the lattice of size
$\xi_\|$
must be treated as a single unit. This gives an obvious explanation to the
fact that the NMR experimental
data~\cite{laplace2012}
was best fitted under the assumption that the doping-introduced disorder
should be averaged over finite-size patches. Our reasoning naturally
equates the size of these patches 
$r_0$
and parameter
$\xi_\|$.

\subsection{The role of the anisotropy}
%%%%%%%%%%%%%%%%%%%%%%%%%%%%%%%%%%%%%%%%%%%%%%%%%%
\label{subsect::anisotropy}
%%%%%%%%%%%%%%%%%%%%%%%%%%%%%%%%%%%%%%%%%%%%%%%%%% 

Evaluating integral in
Eq.~(\ref{eq::integ_eval})
we imposed the following restriction
$\pi^2 \alpha^2 /\beta^2 \ll 1$.
It implies that the Landau functional coefficients should satisfy
${C_\|}/{C_\perp}
\gg
{\pi^2 r_0^2}/{s^2}$.
Since 
BaFe$_2$As$_2$
has two layers per one unit cell, the interlayer distance $s$ equals to
$s = c_0/2 \approx 6.5$\/\AA,
where
$c_0$
is the $c$-axis lattice constant (crystallographic data for
BaFe$_2$As$_2$
may be found in 
Ref.~\onlinecite{data_bafeas2010}).
Using
Eq.~(\ref{eq::r0_estimate}),
we derive
${C_\|}/{C_\perp} \gg 30$.
This means that, for
Eq.~(\ref{eq::integ_final})
to be valid, the Landau functional must be sufficiently anisotropic. It is
not immediately obvious how to estimate the anisotropy of the coefficients
$C_{\perp}$,
$C_{\|}$
for 
BaFe$_2$As$_2$.
Fortunately, the importance of this condition is not too crucial. Indeed,
even in a perfectly isotropic system the
estimate~(\ref{eq::FinalCorrection})
remains valid up to a numerical
factor~\cite{larkin_ovch_inhomogen_sc1972}.

\subsection{Conclusions}

In this paper, we studied the correction to the N\'{e}el temperature
introduced by the inhomogeneities of the doping atoms distribution for
Ru-doped 
BaFe$_2$As$_2$.
Using perturbation theory, we expressed the lowest-order correction to the
N\'{e}el temperature of a Q2D system in terms of the disorder distribution
properties. Previously developed phenomenological model for the disorder in
Ru-doped
BaFe$_2$As$_2$
allows us to complete the calculations. The corrections are found to be
quite small for all doping levels where the material experiences the SDW
transition. This suggests that the N\'{e}el temperature in Ru-doped
BaFe$_2$As$_2$
may be studied using spatially homogeneous models.

\section*{Acknowledgments} 
%This work is partially supported by
%the Russian Foundation for Basic Research (Projects 17-02-00323).
The authors would like to thank Y.~Laplace for useful discussions and help
with interpretations of experimental data.
The authors acknowledge the support by Skoltech NGP Program (Skoltech-MIT
joint project).

\appendix

\begin{widetext}
\section{Dispersion of the second correction $t^{(2)}_{\rm N}$}
%%%%%%%%%%%%%%%%%%%%%%%%%%%%%%%%%%%%%%%%%%%%%%%%%%
\label{app::A}
%%%%%%%%%%%%%%%%%%%%%%%%%%%%%%%%%%%%%%%%%%%%%%%%%% 

In this Appendix we demonstrate that the dispersion of 
$t^{(2)}_{\rm N}$,
given by
Eq.~(\ref{eq::correction0}),
vanishes in the thermodynamic limit. Namely, we intend to prove that the
disorder average of $D$, where
\begin{eqnarray}
D
=
[ t^{(2)}_{\rm N} - \langle t^{(2)}_{\rm N} \rangle ]^2,
\end{eqnarray}
is small for large systems. Since
\begin{eqnarray}
\langle D \rangle
=
\langle [t^{(2)}_{\rm N}]^2 \rangle - \langle t^{(2)}_{\rm N} \rangle^2,
\end{eqnarray}
we need to evaluate
$\langle [t^{(2)}_{\rm N}]^2 \rangle$.
The random quantity 
$[t^{(2)}_{\rm N}]^2$
can be expressed as
\begin{eqnarray}
%%%%%%%%%%%%%%%%%%%%%%%%%%%%%%%%%%%%%%%%%%%%%%%%%% 
\label{t*^2}
%%%%%%%%%%%%%%%%%%%%%%%%%%%%%%%%%%%%%%%%%%%%%%%%%% 
[ t^{(2)}_{\rm N}]^2
=
\frac{1}{V^2}
\int \delta t({\bf r})\delta t({\bf r'}) G({\bf r-r'}) d^3{\bf r}d^3{\bf r'}
\int \delta t({\bf r''})\delta t({\bf r'''}) G({\bf r''-r'''}) d^3{\bf r''}d^3{\bf r'''}.
\end{eqnarray}
In this equation, the Green's function $G$ is given by
Eq.~(\ref{eq::green_f}).

\begin{figure}[t]
	\centering{\includegraphics[width=0.40\textwidth]{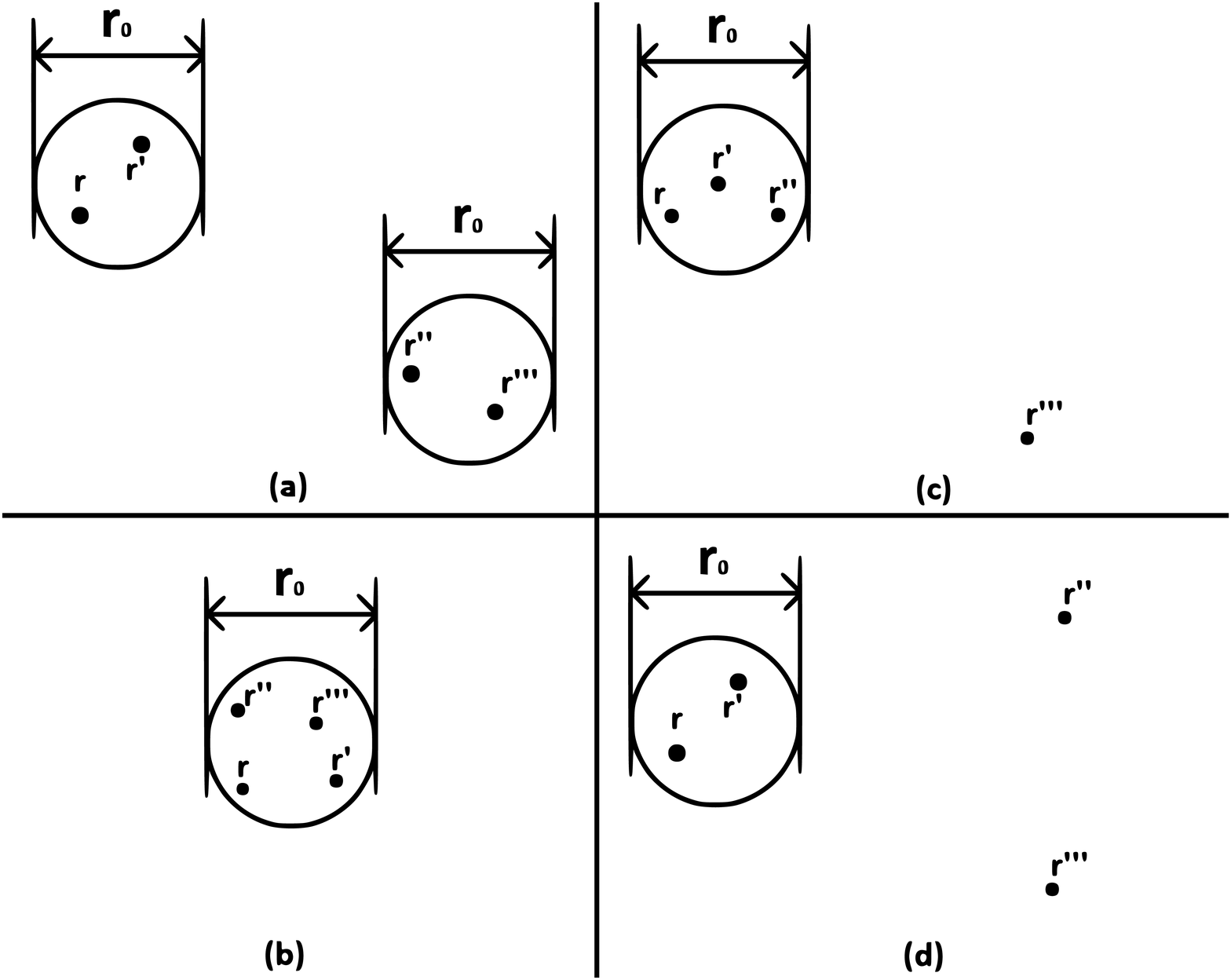}}
	\caption{Illustration to the expansion
Eq.~(\ref{tttt}). 
Locality requires that the average
$\left\langle
	\delta t({\bf r})\delta t({\bf r'})\delta t({\bf r''})\delta t({\bf r'''})
\right\rangle$
vanishes unless even number of points lie within a radius 
$r_0$
from each other. Thus, the configurations shown in panels~(c) and~(d)
correspond to vanishing 
$\langle \delta t\delta t\delta t\delta t \rangle$.
Panel~(a), 
on the other hand, represents the first term on the right-hand side of
Eq.~(\ref{tttt}). 
Two other terms are obtained by permutations of 
${\bf r},\ldots , {\bf r}'''$.
Panel~(b) 
also corresponds to finite average. However, because of the constraint,
requiring all four points be confined within a distance
$\sim r_0$,
after integration over space in
Eq.~(\ref{t*^2}),
one obtains a contribution which, in the thermodynamic limit, is much
smaller than that of panel~(a).
%%%%%%%%%%%%%%%%%%%%%%%%%%%%%%%%%%%%%%%%%%%%%%%%%%
\label{fig1}
%%%%%%%%%%%%%%%%%%%%%%%%%%%%%%%%%%%%%%%%%%%%%%%%%% 
}
\end{figure}

As one can see from
Eq.~(\ref{t*^2}),
to evaluate
$\langle [ t^{(2)}_{\rm N}]^2 \rangle$
we must determine the four-point disorder correlation function
\begin{eqnarray}
\tau^{(4)} ({\bf r}, {\bf r'}, {\bf r''}, {\bf r'''})
=
\langle
\delta t({\bf r})\delta t({\bf r'})\delta t({\bf r''})\delta t({\bf r'''})
\rangle.
\end{eqnarray} 
Below, for simplicity, we will assume that the disorder correlation function is
isotropic. Strictly speaking, this assumption is inapplicable for pnictide compounds, 
and our choice for the disorder correlation function,
Eq.~(\ref{eq::corr_func}),
is explicitly anisotropic. Fortunately, the argumentation presented in this Appendix
can be straightforwardly generalized to anisotropic situations. 

Since the correlations of
$\delta t ({\bf r})$
and
$\delta t ({\bf r}')$
decay if
$| {\bf r} - {\bf r}'| > r_0$,
we can write the following approximate relation
\begin{eqnarray}
%%%%%%%%%%%%%%%%%%%%%%%%%%%%%%%%%%%%%%%%%%%%%%%%%% 
\label{tttt}
%%%%%%%%%%%%%%%%%%%%%%%%%%%%%%%%%%%%%%%%%%%%%%%%%% 
\tau^{(4)} ({\bf r}, {\bf r'}, {\bf r''}, {\bf r'''})
&\approx&
\! \left\langle
	\delta t({\bf r})\delta t({\bf r'})
\right\rangle\!
\left\langle
	\delta t({\bf r''})\delta t({\bf r'''})
\right\rangle
+
\! \left\langle
	\delta t({\bf r})\delta t({\bf r''})
\right\rangle \!
\left\langle
	\delta t({\bf r'})\delta t({\bf r'''})
\right\rangle
+
\! \left\langle
	\delta t({\bf r})\delta t({\bf r'''})
\right\rangle \!
\left\langle 
	\delta t({\bf r'})\delta t({\bf r''})
\right\rangle
\\
\nonumber 
&=&
\tau({\bf r}-{\bf r'})\tau({\bf r''}-{\bf r'''})
+
\tau({\bf r}-{\bf r''})\tau({\bf r'}-{\bf r'''})
+
\tau({\bf r} -{\bf r'''})\tau({\bf r'}-{\bf r''}). 
\end{eqnarray}
which reduces four-point correlation function to the products of
two-point correlation functions $\tau$. This decomposition reminds the Wick
theorem. Yet, justification of
Eq.~(\ref{tttt}) 
is unrelated to the properties of the Gaussian integration, which underpin
the Wick theorem. In essence,
Eq.~(\ref{tttt}) 
assumes that, if point
${\bf r}$
is far from all
${\bf r}'$,
${\bf r}''$,
and
${\bf r}'''$
(see 
Fig.~\ref{fig1}c),
$\delta t ({\bf r})$
is uncorrelated with
$\delta t ({\bf r}')$,
$\delta t ({\bf r}'')$,
and
$\delta t ({\bf r}''')$,
and can be averaged separately from these three. Since
$\langle \delta t ({\bf r})  \rangle = 0$,
the configuration of points shown in
Fig.~\ref{fig1}c
corresponds to vanishing
$\tau^{(4)}$.
Similarly, the configuration of
Fig.~\ref{fig1}d
represents vanishing
$\tau^{(4)}$.

On the other hand, both
Fig.~\ref{fig1}a
and~\ref{fig1}b
depict configurations for which 
$\langle 
\delta t({\bf r})\delta t({\bf r'})\delta t({\bf r''})\delta t({\bf r'''})
\rangle$
is finite. However, in the thermodynamic limit, the configurations of
Fig.~\ref{fig1}a
and~\ref{fig1}b
generate very dissimilar contributions to
$\langle D \rangle$.
Indeed, it is easy to check that the contribution of the
configuration shown in
Fig.~\ref{fig1}b
is smaller by factor of
$r_0^3/V \ll 1$
than the contribution represented by
Fig.~\ref{fig1}a.
Thus, in 
Eq.~(\ref{tttt}),
the configuration of
Fig.~\ref{fig1}b
is justifiably discarded. Among the retained terms, the first one corresponds to
Fig.~\ref{fig1}a.
Two other terms can be obtained by permutations of arguments. 

To simplify calculations it is convenient to introduce new notations:
${\bf r}- {\bf r''} = {\bf R}_1$, ${\bf r'}-{\bf r'''} = {\bf R}_2$,
${\bf r} - {\bf r'} = {\bf R}_3 + {\bf R}_1$,
${\bf r''}-{\bf r'''} = {\bf R}_3 + {\bf R}_2$.
This allows us to re-write
Eq.~(\ref{tttt}) 
\begin{eqnarray}
%%%%%%%%%%%%%%%%%%%%%%%%%%%%%%%%%%%%%%%%%%%%%%%%%% 
\label{tttt2}
%%%%%%%%%%%%%%%%%%%%%%%%%%%%%%%%%%%%%%%%%%%%%%%%%% 
	\! \left\langle \delta t({\bf r})\delta t({\bf r'})\delta t({\bf r''})\delta t({\bf r'''})\right\rangle  \simeq \! \tau( \! {\bf R}_1 \! + {\bf \! R}_3\! )\tau(\! {\bf R}_2 \! + \! {\bf R}_3\! )
	+ \! \tau(\! {\bf R}_1\! )\tau(\! {\bf R}_2\! ) + \! \tau(\! {\bf R}_1 \! + \!{\bf R}_2 \! + \! {\bf R}_3\!)\tau(\!-{\bf R}_3\!). 
\end{eqnarray}
We also define 
\begin{eqnarray}
P\left({\bf R}\right)
=
\int \tau \left( {\bf R'}\right) G\left({\bf R' + R}\right) d^3 {\bf R'}.
%%%%%%%%%%%%%%%%%%%%%%%%%%%%%%%%%%%%%%%%%%%%%%%%%% 
\label{P}
%%%%%%%%%%%%%%%%%%%%%%%%%%%%%%%%%%%%%%%%%%%%%%%%%% 
\end{eqnarray}
In the limit
$|{\bf R}| \rightarrow \infty$
we have
\begin{eqnarray}
%%%%%%%%%%%%%%%%%%%%%%%%%%%%%%%%%%%%%%%%%%%%%%%%%%
\label{eq::P_limit}
%%%%%%%%%%%%%%%%%%%%%%%%%%%%%%%%%%%%%%%%%%%%%%%%%% 
P\left({\bf R}\right) = O(|{\bf R}|^{-1}).
\end{eqnarray}
Combining
Eqs.~(\ref{t*^2})
and~(\ref{tttt2})
with
definition~(\ref{P})
one derives
\begin{eqnarray}	
%%%%%%%%%%%%%%%%%%%%%%%%%%%%%%%%%%%%%%%%%%%%%%%%%% 
\label{t*^2av}
%%%%%%%%%%%%%%%%%%%%%%%%%%%%%%%%%%%%%%%%%%%%%%%%%% 
\langle[ t^{(2)}_{\rm N} ]^2\rangle
=
\frac{1}{V} \int d^3{\bf R}_1 d^3{\bf R}_2 d^3{\bf R}_3
G({\bf R}_1 + {\bf R}_3) G({\bf R}_2 + {\bf R}_3)
	\left[\tau({\bf R}_1+{\bf R}_3)\tau({\bf R}_2+{\bf R}_3) + \tau({\bf R}_1)\tau({\bf R}_2) +
\right.
\\
\nonumber
\left.
\tau({\bf R}_1+{\bf R}_2+{\bf R}_3)\tau(- {\bf R}_3)\right]
=
\left[\langle t^{(2)}_{\rm N}\rangle \right]^2
+ \frac{1}{V}\int  \left[P\left({\bf R}\right)\right]^2 d^3{\bf R}+
\\
\nonumber
+
\frac{1}{V}
\int d^3{\bf R}_1d^3{\bf R}_2 d^3{\bf R}_3 
	\tau({\bf R}_1+{\bf R}_2+{\bf R}_3)\tau(- {\bf R}_3)
	G(\left({\bf R}_1 +{\bf R}_2 + {\bf R}_3\right) - {\bf R}_2) 
	G(-{\bf R}_3 - {\bf R}_2)
=
\\
\nonumber
=
\! \left[ \langle t^{(2)}_{\rm N}\rangle\right]^2 \!
+
\! \frac{1}{V}\!
\int \! \left[\! P\left(\! {\bf R} \! \right)\right]^2
\! d^3{\bf R}
+ \!
\frac{1}{V}
\int \! \left[\! P\left(\! -{\bf R}\! \right)\right]^2 \! d^3{\bf R}
=
\left[\langle t^{(2)}_{\rm N}\rangle\right]^2
+
\frac{2}{V}\int \left[P\left({\bf R}\right)\right]^2 d^3{\bf R}
\end{eqnarray}
	 
If
$V = L^3$,
where $L$ is linear size of the system, then from
Eq.~(\ref{t*^2av})
and~(\ref{eq::P_limit})
it follows
\begin{eqnarray}
\langle D \rangle
= 
\frac{2}{V}
\int \left[P\left({\bf R}\right)\right]^2 d^3{\bf R} 
\sim
\frac{1}{V} \frac{1}{L^2} V = O(L^{-2}).
\end{eqnarray}
If
$L\rightarrow \infty$
then
$\langle D \rangle
\rightarrow 0$.
In other words, the dispersion of
$t_N^{(2)}$
vanishes in the thermodynamic limit.
\end{widetext}

%\bibliographystyle{apsrevlong_no_issn_url}
%\bibliography{inhomogen_tc}

\begin{thebibliography}{21}
\expandafter\ifx\csname natexlab\endcsname\relax\def\natexlab#1{#1}\fi
\expandafter\ifx\csname bibnamefont\endcsname\relax
  \def\bibnamefont#1{#1}\fi
\expandafter\ifx\csname bibfnamefont\endcsname\relax
  \def\bibfnamefont#1{#1}\fi
\expandafter\ifx\csname citenamefont\endcsname\relax
  \def\citenamefont#1{#1}\fi

\bibitem[{\citenamefont{Thaler et~al.}(2010)\citenamefont{Thaler, Ni, Kracher,
  Yan, Bud'ko, and Canfield}}]{data_bafeas2010}
\bibinfo{author}{\bibfnamefont{A.}~\bibnamefont{Thaler}},
  \bibinfo{author}{\bibfnamefont{N.}~\bibnamefont{Ni}},
  \bibinfo{author}{\bibfnamefont{A.}~\bibnamefont{Kracher}},
  \bibinfo{author}{\bibfnamefont{J.~Q.} \bibnamefont{Yan}},
  \bibinfo{author}{\bibfnamefont{S.~L.} \bibnamefont{Bud'ko}},
  \bibnamefont{and} \bibinfo{author}{\bibfnamefont{P.~C.}
  \bibnamefont{Canfield}}, {``}\bibinfo{title}{Physical and magnetic properties
  of
  $\text{Ba}{({\text{Fe}}_{1\ensuremath{-}x}{\text{Ru}}_{x})}_{2}{\text{As}}_{2}$
  single crystals},{''} \bibinfo{journal}{Phys. Rev. B}
  \textbf{\bibinfo{volume}{82}}, \bibinfo{pages}{014534}
  (\bibinfo{year}{2010}).

\bibitem[{\citenamefont{Dhaka et~al.}(2011)\citenamefont{Dhaka, Liu, Fernandes,
  Jiang, Strehlow, Kondo, Thaler, Schmalian, Bud'ko, Canfield
  et~al.}}]{PhysRevLett.107.267002}
\bibinfo{author}{\bibfnamefont{R.~S.} \bibnamefont{Dhaka}},
  \bibinfo{author}{\bibfnamefont{C.}~\bibnamefont{Liu}},
  \bibinfo{author}{\bibfnamefont{R.~M.} \bibnamefont{Fernandes}},
  \bibinfo{author}{\bibfnamefont{R.}~\bibnamefont{Jiang}},
  \bibinfo{author}{\bibfnamefont{C.~P.} \bibnamefont{Strehlow}},
  \bibinfo{author}{\bibfnamefont{T.}~\bibnamefont{Kondo}},
  \bibinfo{author}{\bibfnamefont{A.}~\bibnamefont{Thaler}},
  \bibinfo{author}{\bibfnamefont{J.}~\bibnamefont{Schmalian}},
  \bibinfo{author}{\bibfnamefont{S.~L.} \bibnamefont{Bud'ko}},
  \bibinfo{author}{\bibfnamefont{P.~C.} \bibnamefont{Canfield}},
  \bibnamefont{et~al.}, {``}\bibinfo{title}{What Controls the Phase Diagram and
  Superconductivity in Ru-Substituted
  ${\mathrm{BaFe}}_{2}{\mathrm{As}}_{2}$?},{''} \bibinfo{journal}{Phys. Rev.
  Lett.} \textbf{\bibinfo{volume}{107}}, \bibinfo{pages}{267002}
  (\bibinfo{year}{2011}).

\bibitem[{\citenamefont{Ma et~al.}(2012)\citenamefont{Ma, Ji, Dai, Lu, Eom,
  Kim, Normand, and Yu}}]{ru_doped_sdw_sc_coex_exper2012}
\bibinfo{author}{\bibfnamefont{L.}~\bibnamefont{Ma}},
  \bibinfo{author}{\bibfnamefont{G.~F.} \bibnamefont{Ji}},
  \bibinfo{author}{\bibfnamefont{J.}~\bibnamefont{Dai}},
  \bibinfo{author}{\bibfnamefont{X.~R.} \bibnamefont{Lu}},
  \bibinfo{author}{\bibfnamefont{M.~J.} \bibnamefont{Eom}},
  \bibinfo{author}{\bibfnamefont{J.~S.} \bibnamefont{Kim}},
  \bibinfo{author}{\bibfnamefont{B.}~\bibnamefont{Normand}}, \bibnamefont{and}
  \bibinfo{author}{\bibfnamefont{W.}~\bibnamefont{Yu}},
  {``}\bibinfo{title}{Microscopic Coexistence of Superconductivity and
  Antiferromagnetism in Underdoped
  $\mathrm{Ba}({\mathrm{Fe}}_{1\mathbf{\ensuremath{-}}x}{\mathrm{Ru}}_{x}{)}_{2}{\mathrm{As}}_{2}$},{''}
  \bibinfo{journal}{Phys. Rev. Lett.} \textbf{\bibinfo{volume}{109}},
  \bibinfo{pages}{197002} (\bibinfo{year}{2012}).

\bibitem[{\citenamefont{Zhao et~al.}(2013)\citenamefont{Zhao, Rotundu, Marty,
  Matsuda, Zhao, Setty, Bourret-Courchesne, Hu, and
  Birgeneau}}]{ru_doped_neutron_resonant_mode2013}
\bibinfo{author}{\bibfnamefont{J.}~\bibnamefont{Zhao}},
  \bibinfo{author}{\bibfnamefont{C.~R.} \bibnamefont{Rotundu}},
  \bibinfo{author}{\bibfnamefont{K.}~\bibnamefont{Marty}},
  \bibinfo{author}{\bibfnamefont{M.}~\bibnamefont{Matsuda}},
  \bibinfo{author}{\bibfnamefont{Y.}~\bibnamefont{Zhao}},
  \bibinfo{author}{\bibfnamefont{C.}~\bibnamefont{Setty}},
  \bibinfo{author}{\bibfnamefont{E.}~\bibnamefont{Bourret-Courchesne}},
  \bibinfo{author}{\bibfnamefont{J.}~\bibnamefont{Hu}}, \bibnamefont{and}
  \bibinfo{author}{\bibfnamefont{R.~J.} \bibnamefont{Birgeneau}},
  {``}\bibinfo{title}{Effect of Electron Correlations on Magnetic Excitations
  in the Isovalently Doped Iron-Based Superconductor
  $\mathrm{Ba}({\mathrm{Fe}}_{1\mathbf{\ensuremath{-}}x}{\mathrm{Ru}}_{x}{)}_{2}{\mathrm{As}}_{2}$},{''}
  \bibinfo{journal}{Phys. Rev. Lett.} \textbf{\bibinfo{volume}{110}},
  \bibinfo{pages}{147003} (\bibinfo{year}{2013}).

\bibitem[{\citenamefont{Wang et~al.}(2013)\citenamefont{Wang, Berlijn, Wang,
  Lin, Hirschfeld, and Ku}}]{first_princip_ru_doped2013}
\bibinfo{author}{\bibfnamefont{L.}~\bibnamefont{Wang}},
  \bibinfo{author}{\bibfnamefont{T.}~\bibnamefont{Berlijn}},
  \bibinfo{author}{\bibfnamefont{Y.}~\bibnamefont{Wang}},
  \bibinfo{author}{\bibfnamefont{C.-H.} \bibnamefont{Lin}},
  \bibinfo{author}{\bibfnamefont{P.~J.} \bibnamefont{Hirschfeld}},
  \bibnamefont{and} \bibinfo{author}{\bibfnamefont{W.}~\bibnamefont{Ku}},
  {``}\bibinfo{title}{Effects of Disordered Ru Substitution in
  ${\mathrm{BaFe}}_{2}{\mathrm{As}}_{2}$: Possible Realization of
  Superdiffusion in Real Materials},{''} \bibinfo{journal}{Phys. Rev. Lett.}
  \textbf{\bibinfo{volume}{110}}, \bibinfo{pages}{037001}
  (\bibinfo{year}{2013}).

\bibitem[{\citenamefont{Ye et~al.}(2014)\citenamefont{Ye, Zhang, Chen, Xu,
  Jiang, Niu, Wen, Xing, Wang, Jin et~al.}}]{iron_based2014}
\bibinfo{author}{\bibfnamefont{Z.~R.} \bibnamefont{Ye}},
  \bibinfo{author}{\bibfnamefont{Y.}~\bibnamefont{Zhang}},
  \bibinfo{author}{\bibfnamefont{F.}~\bibnamefont{Chen}},
  \bibinfo{author}{\bibfnamefont{M.}~\bibnamefont{Xu}},
  \bibinfo{author}{\bibfnamefont{J.}~\bibnamefont{Jiang}},
  \bibinfo{author}{\bibfnamefont{X.~H.} \bibnamefont{Niu}},
  \bibinfo{author}{\bibfnamefont{C.~H.~P.} \bibnamefont{Wen}},
  \bibinfo{author}{\bibfnamefont{L.~Y.} \bibnamefont{Xing}},
  \bibinfo{author}{\bibfnamefont{X.~C.} \bibnamefont{Wang}},
  \bibinfo{author}{\bibfnamefont{C.~Q.} \bibnamefont{Jin}},
  \bibnamefont{et~al.}, {``}\bibinfo{title}{Extraordinary Doping Effects on
  Quasiparticle Scattering and Bandwidth in Iron-Based Superconductors},{''}
  \bibinfo{journal}{Phys. Rev. X} \textbf{\bibinfo{volume}{4}},
  \bibinfo{pages}{031041} (\bibinfo{year}{2014}).

\bibitem[{\citenamefont{Reticcioli et~al.}(2017)\citenamefont{Reticcioli,
  Profeta, Franchini, and Continenza}}]{dft_ru_doped2017}
\bibinfo{author}{\bibfnamefont{M.}~\bibnamefont{Reticcioli}},
  \bibinfo{author}{\bibfnamefont{G.}~\bibnamefont{Profeta}},
  \bibinfo{author}{\bibfnamefont{C.}~\bibnamefont{Franchini}},
  \bibnamefont{and}
  \bibinfo{author}{\bibfnamefont{A.}~\bibnamefont{Continenza}},
  {``}\bibinfo{title}{Ru doping in iron-based pnictides: The ``unfolded''
  dominant role of structural effects for superconductivity},{''}
  \bibinfo{journal}{Phys. Rev. B} \textbf{\bibinfo{volume}{95}},
  \bibinfo{pages}{214510} (\bibinfo{year}{2017}).

\bibitem[{\citenamefont{Prozorov et~al.}(2014)\citenamefont{Prozorov,
  Ko\ifmmode~\acute{n}\else \'{n}\fi{}czykowski, Tanatar, Thaler, Bud'ko,
  Canfield, Mishra, and Hirschfeld}}]{irradiation_disorder2014}
\bibinfo{author}{\bibfnamefont{R.}~\bibnamefont{Prozorov}},
  \bibinfo{author}{\bibfnamefont{M.}~\bibnamefont{Ko\ifmmode~\acute{n}\else
  \'{n}\fi{}czykowski}}, \bibinfo{author}{\bibfnamefont{M.~A.}
  \bibnamefont{Tanatar}},
  \bibinfo{author}{\bibfnamefont{A.}~\bibnamefont{Thaler}},
  \bibinfo{author}{\bibfnamefont{S.~L.} \bibnamefont{Bud'ko}},
  \bibinfo{author}{\bibfnamefont{P.~C.} \bibnamefont{Canfield}},
  \bibinfo{author}{\bibfnamefont{V.}~\bibnamefont{Mishra}}, \bibnamefont{and}
  \bibinfo{author}{\bibfnamefont{P.~J.} \bibnamefont{Hirschfeld}},
  {``}\bibinfo{title}{Effect of Electron Irradiation on Superconductivity in
  Single Crystals of
  $\mathrm{Ba}({\mathrm{Fe}}_{1\ensuremath{-}x}{\mathrm{Ru}}_{x}{)}_{2}{\mathrm{As}}_{2}$
  ($x=0.24$)},{''} \bibinfo{journal}{Phys. Rev. X}
  \textbf{\bibinfo{volume}{4}}, \bibinfo{pages}{041032} (\bibinfo{year}{2014}).

\bibitem[{\citenamefont{Park et~al.}(2009)\citenamefont{Park, Inosov,
  Niedermayer, Sun, Haug, Christensen, Dinnebier, Boris, Drew, Schulz
  et~al.}}]{PSexp1}
\bibinfo{author}{\bibfnamefont{J.~T.} \bibnamefont{Park}},
  \bibinfo{author}{\bibfnamefont{D.~S.} \bibnamefont{Inosov}},
  \bibinfo{author}{\bibfnamefont{C.}~\bibnamefont{Niedermayer}},
  \bibinfo{author}{\bibfnamefont{G.~L.} \bibnamefont{Sun}},
  \bibinfo{author}{\bibfnamefont{D.}~\bibnamefont{Haug}},
  \bibinfo{author}{\bibfnamefont{N.~B.} \bibnamefont{Christensen}},
  \bibinfo{author}{\bibfnamefont{R.}~\bibnamefont{Dinnebier}},
  \bibinfo{author}{\bibfnamefont{A.~V.} \bibnamefont{Boris}},
  \bibinfo{author}{\bibfnamefont{A.~J.} \bibnamefont{Drew}},
  \bibinfo{author}{\bibfnamefont{L.}~\bibnamefont{Schulz}},
  \bibnamefont{et~al.}, {``}\bibinfo{title}{Electronic Phase Separation in the
  Slightly Underdoped Iron Pnictide Superconductor
  ${\mathrm{Ba}}_{1-x}{\mathrm{K}}_{x}{\mathrm{Fe}}_{2}{\mathrm{As}}_{2}$},{''}
  \bibinfo{journal}{Phys. Rev. Lett.} \textbf{\bibinfo{volume}{102}},
  \bibinfo{pages}{117006} (\bibinfo{year}{2009}).

\bibitem[{\citenamefont{Inosov et~al.}(2009)\citenamefont{Inosov, Leineweber,
  Yang, Park, Christensen, Dinnebier, Sun, Niedermayer, Haug, Stephens
  et~al.}}]{PSexp2}
\bibinfo{author}{\bibfnamefont{D.~S.} \bibnamefont{Inosov}},
  \bibinfo{author}{\bibfnamefont{A.}~\bibnamefont{Leineweber}},
  \bibinfo{author}{\bibfnamefont{X.}~\bibnamefont{Yang}},
  \bibinfo{author}{\bibfnamefont{J.~T.} \bibnamefont{Park}},
  \bibinfo{author}{\bibfnamefont{N.~B.} \bibnamefont{Christensen}},
  \bibinfo{author}{\bibfnamefont{R.}~\bibnamefont{Dinnebier}},
  \bibinfo{author}{\bibfnamefont{G.~L.} \bibnamefont{Sun}},
  \bibinfo{author}{\bibfnamefont{C.}~\bibnamefont{Niedermayer}},
  \bibinfo{author}{\bibfnamefont{D.}~\bibnamefont{Haug}},
  \bibinfo{author}{\bibfnamefont{P.~W.} \bibnamefont{Stephens}},
  \bibnamefont{et~al.}, {``}\bibinfo{title}{Suppression of the structural phase
  transition and lattice softening in slightly underdoped
  ${\text{Ba}}_{1\ensuremath{-}x}{\text{K}}_{x}{\text{Fe}}_{2}{\text{As}}_{2}$
  with electronic phase separation},{''} \bibinfo{journal}{Phys. Rev. B}
  \textbf{\bibinfo{volume}{79}}, \bibinfo{pages}{224503}
  (\bibinfo{year}{2009}).

\bibitem[{\citenamefont{Lang et~al.}(2010)\citenamefont{Lang, Grafe, Paar,
  Hammerath, Manthey, Behr, Werner, and B\"uchner}}]{PSexp3}
\bibinfo{author}{\bibfnamefont{G.}~\bibnamefont{Lang}},
  \bibinfo{author}{\bibfnamefont{H.-J.} \bibnamefont{Grafe}},
  \bibinfo{author}{\bibfnamefont{D.}~\bibnamefont{Paar}},
  \bibinfo{author}{\bibfnamefont{F.}~\bibnamefont{Hammerath}},
  \bibinfo{author}{\bibfnamefont{K.}~\bibnamefont{Manthey}},
  \bibinfo{author}{\bibfnamefont{G.}~\bibnamefont{Behr}},
  \bibinfo{author}{\bibfnamefont{J.}~\bibnamefont{Werner}}, \bibnamefont{and}
  \bibinfo{author}{\bibfnamefont{B.}~\bibnamefont{B\"uchner}},
  {``}\bibinfo{title}{Nanoscale Electronic Order in Iron Pnictides},{''}
  \bibinfo{journal}{Phys. Rev. Lett.} \textbf{\bibinfo{volume}{104}},
  \bibinfo{pages}{097001} (\bibinfo{year}{2010}).

\bibitem[{\citenamefont{Bonville et~al.}(2010)\citenamefont{Bonville,
  Rullier-Albenque, Colson, and Forget}}]{phasep_bafeas_exp2010}
\bibinfo{author}{\bibfnamefont{P.}~\bibnamefont{Bonville}},
  \bibinfo{author}{\bibfnamefont{F.}~\bibnamefont{Rullier-Albenque}},
  \bibinfo{author}{\bibfnamefont{D.}~\bibnamefont{Colson}}, \bibnamefont{and}
  \bibinfo{author}{\bibfnamefont{A.}~\bibnamefont{Forget}},
  {``}\bibinfo{title}{Incommensurate spin density wave in Co-doped
  BaFe{$_2$}As{$_2$}},{''} \bibinfo{journal}{EPL}
  \textbf{\bibinfo{volume}{89}}, \bibinfo{pages}{67008} (\bibinfo{year}{2010}).

\bibitem[{\citenamefont{Shen et~al.}(2011)\citenamefont{Shen, Zeng, Chen, He,
  Wang, Yang, and Wen}}]{epl_inhomogen_sc_2011}
\bibinfo{author}{\bibfnamefont{B.}~\bibnamefont{Shen}},
  \bibinfo{author}{\bibfnamefont{B.}~\bibnamefont{Zeng}},
  \bibinfo{author}{\bibfnamefont{G.~F.} \bibnamefont{Chen}},
  \bibinfo{author}{\bibfnamefont{J.~B.} \bibnamefont{He}},
  \bibinfo{author}{\bibfnamefont{D.~M.} \bibnamefont{Wang}},
  \bibinfo{author}{\bibfnamefont{H.}~\bibnamefont{Yang}}, \bibnamefont{and}
  \bibinfo{author}{\bibfnamefont{H.~H.} \bibnamefont{Wen}},
  {``}\bibinfo{title}{Intrinsic percolative superconductivity in
  K$_x$Fe$_{2-y}$Se$_2$ single crystals},{''} \bibinfo{journal}{EPL}
  \textbf{\bibinfo{volume}{96}}, \bibinfo{pages}{37010} (\bibinfo{year}{2011}).

\bibitem[{\citenamefont{Bernhard et~al.}(2012)\citenamefont{Bernhard, Wang,
  Nuccio, Schulz, Zaharko, Larsen, Aristizabal, Willis, Drew, Varma
  et~al.}}]{phasep_exp2012}
\bibinfo{author}{\bibfnamefont{C.}~\bibnamefont{Bernhard}},
  \bibinfo{author}{\bibfnamefont{C.~N.} \bibnamefont{Wang}},
  \bibinfo{author}{\bibfnamefont{L.}~\bibnamefont{Nuccio}},
  \bibinfo{author}{\bibfnamefont{L.}~\bibnamefont{Schulz}},
  \bibinfo{author}{\bibfnamefont{O.}~\bibnamefont{Zaharko}},
  \bibinfo{author}{\bibfnamefont{J.}~\bibnamefont{Larsen}},
  \bibinfo{author}{\bibfnamefont{C.}~\bibnamefont{Aristizabal}},
  \bibinfo{author}{\bibfnamefont{M.}~\bibnamefont{Willis}},
  \bibinfo{author}{\bibfnamefont{A.~J.} \bibnamefont{Drew}},
  \bibinfo{author}{\bibfnamefont{G.~D.} \bibnamefont{Varma}},
  \bibnamefont{et~al.}, {``}\bibinfo{title}{Muon spin rotation study of
  magnetism and superconductivity in
  Ba(Fe${}_{1\ensuremath{-}x}$Co${}_{x}$)${}_{2}$As${}_{2}$ single
  crystals},{''} \bibinfo{journal}{Phys. Rev. B} \textbf{\bibinfo{volume}{86}},
  \bibinfo{pages}{184509} (\bibinfo{year}{2012}).

\bibitem[{\citenamefont{Civardi et~al.}(2016)\citenamefont{Civardi, Moroni,
  Babij, Bukowski, and Carretta}}]{phasep_exp2016}
\bibinfo{author}{\bibfnamefont{E.}~\bibnamefont{Civardi}},
  \bibinfo{author}{\bibfnamefont{M.}~\bibnamefont{Moroni}},
  \bibinfo{author}{\bibfnamefont{M.}~\bibnamefont{Babij}},
  \bibinfo{author}{\bibfnamefont{Z.}~\bibnamefont{Bukowski}}, \bibnamefont{and}
  \bibinfo{author}{\bibfnamefont{P.}~\bibnamefont{Carretta}},
  {``}\bibinfo{title}{Superconductivity Emerging from an Electronic Phase
  Separation in the Charge Ordered Phase of
  ${\mathrm{RbFe}}_{2}{\mathrm{As}}_{2}$},{''} \bibinfo{journal}{Phys. Rev.
  Lett.} \textbf{\bibinfo{volume}{117}}, \bibinfo{pages}{217001}
  (\bibinfo{year}{2016}).

\bibitem[{\citenamefont{Sboychakov et~al.}(2013)\citenamefont{Sboychakov,
  Rozhkov, Kugel, Rakhmanov, and Nori}}]{phasep_pnics_ours2013}
\bibinfo{author}{\bibfnamefont{A.~O.} \bibnamefont{Sboychakov}},
  \bibinfo{author}{\bibfnamefont{A.~V.} \bibnamefont{Rozhkov}},
  \bibinfo{author}{\bibfnamefont{K.~I.} \bibnamefont{Kugel}},
  \bibinfo{author}{\bibfnamefont{A.~L.} \bibnamefont{Rakhmanov}},
  \bibnamefont{and} \bibinfo{author}{\bibfnamefont{F.}~\bibnamefont{Nori}},
  {``}\bibinfo{title}{Electronic phase separation in iron pnictides},{''}
  \bibinfo{journal}{Phys. Rev. B} \textbf{\bibinfo{volume}{88}},
  \bibinfo{pages}{195142} (\bibinfo{year}{2013}).

\bibitem[{\citenamefont{de' Medici}(2017)}]{hund_phasep_theor2017}
\bibinfo{author}{\bibfnamefont{L.}~\bibnamefont{de' Medici}},
  {``}\bibinfo{title}{Hund's Induced Fermi-Liquid Instabilities and Enhanced
  Quasiparticle Interactions},{''} \bibinfo{journal}{Phys. Rev. Lett.}
  \textbf{\bibinfo{volume}{118}}, \bibinfo{pages}{167003}
  (\bibinfo{year}{2017}).

\bibitem[{\citenamefont{Laplace et~al.}(2012)\citenamefont{Laplace, Bobroff,
  Brouet, Collin, Rullier-Albenque, Colson, and Forget}}]{laplace2012}
\bibinfo{author}{\bibfnamefont{Y.}~\bibnamefont{Laplace}},
  \bibinfo{author}{\bibfnamefont{J.}~\bibnamefont{Bobroff}},
  \bibinfo{author}{\bibfnamefont{V.}~\bibnamefont{Brouet}},
  \bibinfo{author}{\bibfnamefont{G.}~\bibnamefont{Collin}},
  \bibinfo{author}{\bibfnamefont{F.}~\bibnamefont{Rullier-Albenque}},
  \bibinfo{author}{\bibfnamefont{D.}~\bibnamefont{Colson}}, \bibnamefont{and}
  \bibinfo{author}{\bibfnamefont{A.}~\bibnamefont{Forget}},
  {``}\bibinfo{title}{Nanoscale-textured superconductivity in Ru-substituted
  BaFe${}_{2}$As${}_{2}$: A challenge to a universal phase diagram for the
  pnictides},{''} \bibinfo{journal}{Phys. Rev. B}
  \textbf{\bibinfo{volume}{86}}, \bibinfo{pages}{020510}
  (\bibinfo{year}{2012}).

\bibitem[{\citenamefont{Larkin and
  Ovchinnikov}(1972)}]{larkin_ovch_inhomogen_sc1972}
\bibinfo{author}{\bibfnamefont{A.~I.} \bibnamefont{Larkin}} \bibnamefont{and}
  \bibinfo{author}{\bibfnamefont{Y.~N.} \bibnamefont{Ovchinnikov}},
  {``}\bibinfo{title}{Influence of inhomogeneities on superconductors
  properties},{''} \bibinfo{journal}{Sov. Phys. JETP}
  \textbf{\bibinfo{volume}{34}}, \bibinfo{pages}{651} (\bibinfo{year}{1972}).

\bibitem[{\citenamefont{de~Gennes}(1996)}]{deGennes}
\bibinfo{author}{\bibfnamefont{P.}~\bibnamefont{de~Gennes}},
  \emph{\bibinfo{title}{Superconductivity of Metals and Alloys}}
  (\bibinfo{publisher}{Addison-Wesley}, \bibinfo{address}{Reading,
  Massachusetts}, \bibinfo{year}{1996}).

\bibitem[{\citenamefont{White and Geballe}(1984)}]{geballe1984}
\bibinfo{author}{\bibfnamefont{R.}~\bibnamefont{White}} \bibnamefont{and}
  \bibinfo{author}{\bibfnamefont{T.}~\bibnamefont{Geballe}},
  \emph{\bibinfo{title}{Long Range Order in Solids}}
  (\bibinfo{publisher}{Academic, New York and London}, \bibinfo{year}{1984}).

\end{thebibliography}

\end{document}